
\input harvmac

\def\ao{A_{\rm open}}
\Title{PUPT--1283, HUTP--91/A047}
{{\vbox {\centerline{Open String Theory in 1+1 Dimensions}
\bigskip
}}}

\centerline{M. Bershadsky}
\medskip\centerline
{\it Lyman Laboratory,}
\centerline{\it Harvard University,}
\centerline{\it Cambridge, MA 02138}
\bigskip
\centerline{and}
\bigskip
\centerline{D. Kutasov}
\medskip\centerline
{\it Joseph Henry Laboratories,}
\centerline{\it Princeton University,}
\centerline{\it Princeton, NJ 08544.}

\vskip .2in

\noindent
We show that tree level open two dimensional string theory
is exactly solvable; the solution exhibits some unusual features,
and is qualitatively different from the closed case.
The open string ``tachyon'' S -- matrix
describes free fermions, which can be interpreted as the quarks
at the ends of the string. These ``quarks'' live naturally
on a lattice in space-time. We also find
an exact vacuum solution of the theory, corresponding to
a charged black hole.

\Date{9/91}

The study of string propagation in two dimensional (2d) space-time may
provide us with useful toy models for critical string theory and 2d quantum
gravity. String dynamics in such backgrounds is expected to be exactly
solvable; this phenomenon is closely related to the extremely small
number of degrees of freedom in these vacua, and is probably not shared
by higher dimensional string theories compactified to two
dimensions. The models most intensely studied to date are those
of closed bosonic strings in 2d\foot{The $N=1$ supersymmetric
extension was studied in
\ref\KUS{D. Kutasov and N. Seiberg, Phys. Lett. {\bf 251B} (1990) 67;
D. Kutasov, Princeton preprint PUPT-1277 (1991).}
\ref\DIIFK{P. Di Francesco and D. Kutasov, Princeton preprint
PUPT-1276 (1991).}; $N=2$ strings were discussed in
\ref\Ntwo{H. Ooguri and C. Vafa, Mod. Phys. Lett. {\bf A5} (1990) 1389;
Nucl. Phys. {\bf B361} (1991) 469.}.}.
Using large $N$ matrix model techniques
\ref\SURF{V. Kazakov, Phys. Lett. {\bf 150B} (1985) 282; F. David, Nucl.
Phys. {\bf B257} (1985) 45; J. Ambjorn, B. Durhuus and J. Frohlich,
Nucl. Phys. {\bf B257} (1985) 433; V. Kazakov, I. Kostov and A. Migdal,
Phys. Lett. {\bf 157B} (1985) 295.},
\ref\ORIG{D. Gross and A. Migdal, Phys. Rev. Lett. {\bf 64}
(1990) 127; M. Douglas and S. Shenker, Nucl. Phys. {\bf B335} (1990) 635;
E. Brezin and V. Kazakov, Phys. Lett. {\bf 236B} (1990) 144.},
\ref\CONE{
E. Brezin, V. Kazakov, and Al. Zamolodchikov, Nucl. Phys. {\bf B338}
(1990) 673; D. Gross and N. Miljkovi\'c, Phys. Lett {\bf 238B} (1990) 217;
P. Ginsparg and J. Zinn-Justin, Phys. Lett {\bf 240B} (1990) 333;
G. Parisi, Phys. Lett. {\bf 238B} (1990) 209.},
these were shown to be equivalent to theories of free fermions; their
discrete formulation is quite well understood. The continuum path integral
(Liouville) approach
\ref\POL{A. Polyakov, Phys. Lett. {\bf B103} (1981) 207, 211.}
is less developed.
It is
known
\ref\PO{A. Polyakov, Mod. Phys. Lett. {\bf A6} (1991) 635.},
\ref\DFK{P. Di Francesco and D. Kutasov, Phys. Lett.
{\bf261B} (1991) 385.}
that the continuum correlation functions have
an extremely simple form. In \DIIFK\
it was shown that the underlying reason for this simplicity is a partial
decoupling of a certain infinite set of ``discrete'' states. This observation
will play a role below.

In this letter we will study open string theory in two dimensional space-time.
We will find that the tree level dynamics is quite different
from the closed sector. On the one hand, the S -- matrix
which is again exactly calculable, exhibits a more complicated
pole structure, and has some features which resemble higher
dimensional string theories; on the other, this S -- matrix
will be seen to follow from a field theory of massless free fermions
with a lattice propagator!

The theory of world sheet gravity on a manifold $\cal M$ with the
topology of a disk
\ref\open{C. Callan, C. Lovelace, C. Nappi and S. Yost, Nucl. Phys.
{\bf B288} (1987) 525; R. Metsaev, A. Tseytlin, Nucl. Phys. {\bf B298}
(1988) 109.}, corresponds
to 2d string theory with open and closed strings.
The action is:
\eqn\action{\eqalign{
{\cal S}=&{1\over2\pi}\int_{\cal M}d^2\xi\sqrt{g}\left[g^{ab}G_{\mu\nu}
(X)\partial_a X^\mu\partial_b X^\nu-R^{(2)}\Phi(X)+T(X)\right]\cr
+&{1\over\pi}\int_{\partial\cal M}d\xi\left[A_\mu(X)\partial_\xi X^\mu
-K\Phi(X)+g^{1\over4}T_B(X)\right]\cr}
}
where $g_{ab}$ is the world sheet metric, $R^{(2)}$ its scalar
curvature, $g={\rm det} g_{ab}$,
and $K$ the extrinsic curvature on the boundary $\partial\cal M$.
$X^\mu$, $\mu=0,1$ parametrize the space-time manifold, and we have
introduced couplings to a general space-time metric $G_{\mu\nu}$,
gauge field $A_\mu$, dilaton $\Phi$, and bulk and boundary ``tachyon''
fields $T, T_B$. For $T=T_B=0$, the space-time equations of motion
for slowly
varying fields are \open:
\eqn\eqmot{\eqalign{
R_{\mu\nu}=&\nabla_\mu\nabla_\nu\Phi\cr
(\nabla^\nu F_\mu^\lambda)[G-F^2]^{-1}_{\lambda\nu}+&{1\over2}(\nabla^\nu
\Phi) F_{\mu\nu}=0\cr}}
We will start by considering the simplest (exact)
solution, $F_{\mu\nu}=R_{\mu\nu}
=0$ and $\Phi=Q X^0-2i\alpha_0  X^1$. As usual \POL\ one should
also fix the area of the surface ${\cal M}$ and the length of the
boundary $\partial\cal M$ or,
Laplace transforming, turn on condensates of $T(X)=\mu e^{\alpha_+ X^0}$
and $T_B=\rho e^{{1\over2}\alpha_+ X^0}$.
The parameters $Q, \alpha_+$ are related by gauge invariance
to $\alpha_0$
\ref\DDK{F. David, Mod. Phys. Lett. {\bf A3} (1988) 1651;
J. Distler and H. Kawai, Nucl. Phys. {\bf B321} (1989) 509.}:
$$-Q=\alpha_++\alpha_-;\;2\alpha_0=\alpha_--\alpha_+;\;\alpha_+
\alpha_-=2$$
where $|\alpha_+|<|\alpha_-|$. The fields $X^0$, $X^1$ satisfy
Neumann boundary conditions $\partial_n X^1=0$,
$\partial_n X^0=\rho e^{{1\over2}\alpha_+X^0}$.
The boundary conditions usually considered in the matrix models
\ref\mms{E. Martinec, G. Moore and N. Seiberg, Phys. Lett. {\bf 263B} (1991)
190.},
\ref\moor{G.
Moore, N. Seiberg, Rutgers preprint RU-91-29, (1991).}
are Dirichlet boundary conditions (for $X^1$). Therefore their results can not
be directly borrowed to our situation (which is modeled after critical
open string theory).

The only field theoretic degree of freedom in this theory, in the
open string sector, is the massless tachyon field $T_B$, with
on shell vertex operator:
\eqn\vertex{T_k^{(\pm)}=\int_{\partial\cal M}d\xi e^{i{k\over2} X^1
+{\beta_\pm\over2}X^0};\; \beta_\pm=-{Q\over2}\pm (k-\alpha_0)}
$T_k^{(+)}$ ($T_k^{(-)}$) correspond to right (left) moving tachyons
in space-time. In addition, the spectrum includes oscillator states
at discrete values of the momenta $\sqrt{2}k\in Z$.

As in closed string theory, for finite bulk cosmological constant $\mu$ and
boundary cosmological constant $\rho$, the generic amplitudes
$\ao(k_1,...,k_n)=\langle T_{k_1}...T_{k_N}\rangle$ are not known.
However, the bulk amplitudes (see e.g. \DIIFK\ for a discussion), which satisfy
$\sum_{i=1}^N\beta_i=-Q$ (in addition to the obvious
$\sum_{i=1}^N k_i=2\alpha_0$) are calculable
\ref\GTW{ A. Gupta, S. Trivedi and M. Wise, Nucl. Phys. {\bf B340}
(1990) 475.}.
By performing the $X^0$ zero mode integral, one finds
in that case:
\eqn\tachcor{\eqalign{
\ao(k_1,...,k_N)=&\log\Lambda
\int_1^\infty d\xi_{N-1}
\int_1^{\xi_{N-1}}d\xi_{N-2}...\int_1^{\xi_4}d\xi_3\cr
&\langle T_{k_1}(0)
T_{k_2}(1) T_{k_3}(\xi_3)...T_{k_{N-1}}(\xi_{N-1})T_{k_N}(\infty)\rangle\cr}}
where $\Lambda$ is a UV cutoff, and
the correlator $\langle T_1...T_N\rangle$ is evaluated
using the free field propagators $\langle X^\mu(\xi)X^\nu(0)\rangle=-4
\delta^{\mu\nu}\log|\xi|$; the boundary ``cosmological'' interaction
in \action\ has disappeared. The factor of $\log\Lambda$ signals a bulk
effect, being the effective volume of $X^0$ (there is also a factor
of the volume of $X^1$ which we didn't write).
Note that, as usual, the basic object in the open sector
is the amplitude with a given ordering of the tachyons (up to
cyclic permutations)
--$SL(2, R)$ transformations do not permute the particles.
Different orderings correspond to different channels.
Of course, in the
end one has to sum over all non cyclic permutations to recover the total
amplitude.
All we have said so far is valid
(or can be trivially generalized) for any dimension of space-time. The magic
of 2d string theory allows one to explicitly evaluate the integrals \tachcor,
due to simplifications in the dynamics. We will next present
the results and discuss them; the derivation and further discussion will appear
elsewhere.

It is clear from \vertex\ and the behavior of closed 2d string amplitudes,
that the chirality of the massless ``tachyon'' \vertex\ should play
a central role. The generic tachyon amplitude \tachcor\ has the form:
\eqn\chir{\ao(k_1,...,k_N)=\langle(T^{(+)})^{n_1}
(T^{(-)})^{m_1}(T^{(+)})^{n_2}(T^{(-)})^{m_2}...(T^{(+)})^{n_r}
(T^{(-)})^{m_r}\rangle}
By \chir\ we mean the (ordered) amplitude consisting of $n_1$ right moving
tachyons, followed by $m_1$ left moving ones, etc. By cyclic permutation
symmetry we can always assure that $n_1, m_r\not=0$, and this is how we define
\chir.
The first interesting property
of the amplitudes is that $\ao(k_i)=0$ if $r>1$.
In other words, only amplitudes of the form:
\eqn\Anm{\ao^{(n,m)}(k_1,...,k_n, p_1,...,p_m)=
\langle T_{k_1}^{(+)}...T_{k_n}^{(+)}T_{p_1}^{(-)}...T_{p_m}^{(-)}
\rangle}
can be non zero (for generic $k,p$).
This is the first hint of an almost complete separation of the
dynamics of left and right moving tachyons.
Recall that in the closed sector only $A_{\rm closed}^{
(n,1)}, A_{\rm closed}^{(1,n)}\not=0$. The structure
of $\ao^{(n,m)}$ \Anm\ is much more complex than in \DFK;
it is very convenient to parametrize it in terms of the natural variables
for \vertex, $m_i={1\over2}\beta_i^2-{1\over2}k_i^2$.
The amplitudes \Anm\ are most simply expressed in terms of
the functions $F_n$:
\eqn\fun{F_n(k_1,k_2,...,k_n)=\prod_{l=1}^{n-1}{1\over\sin\pi\sum_{i=1}^l
m_i}}
and are given by:
\eqn\nm{\ao^{(n,m)}(k_1,...k_n,p_1,...,p_m)=
\left[\prod_{i=1}^{n+m}{1\over\Gamma(1-m_i)}\right]
F_n(k_1,...,k_n)F_m(p_1,...,p_m)}
with the kinematic constraints
$\sum_{i=1} ^n m(k_i)=2-m~,~~\sum_{i=1} ^m m(p_i)=2-n$.
The form \nm\ deserves a few comments:

\noindent{}1) We see that in the 2d open string, the pole structure
is much more intricate than in the closed one, where $A^{(n,1)}$ is given
by a form similar to \nm\ with $F_nF_m$ replaced by
$\prod_{i=1}^n\Gamma(m_i)$ (and the rest of $A^{(n,m)}$ vanish).
In particular in the open string case there are many ``moving'' (codimension
1) poles, which are absent in the closed case.

\noindent{}2) It is interesting to follow the origin of different
poles by factorization. The poles appear when several neighbouring
particles collide.
We see by \fun, \nm, that they occur when
$\sum_{i=1}^lm_i=r\in Z$ ($m_i$ is either $m(k_i)$ or $m(p_i)$;
$l=1,2,...,n$ (or $m$)).
Poles in the variable $\sum_{i=1}^l m_i$ come from two different regions
of moduli space of \tachcor. The first is from the region where
$T_1,...,T_l$ approach each other. It is easy to see that this gives
poles at $\sum_{i=1}^lm_i=L=1,2,3,...$, corresponding to intermediate
states at (oscillator) level $(L-1)(l-1)$. Only these levels appear
in intermediate states due to kinematic restrictions (see \DIIFK).
The second source of poles is from the region where (say)
$T_{k_1}^{(+)},...,T_{k_l}^{(+)}$ approach
$T_{p_1}^{(-)},...,T_{p_r}^{(-)}$, a subset of the tachyons of the opposite
chirality adjacent to the $T_{k_i}^{(+)}$. Here we have the following
situation: if $r<m$ (i.e. the negative chirality tachyons involved are only
a subset of all $T^{(-)}$), only tachyon poles are possible, since higher
poles would correspond to oscillator states at generic momenta, which must
by general arguments be BRST exact. The tachyon poles in this channel occur
at $\sum_{i=1}^lm_i=1-r=0,-1,-2,...,1-m$. This leaves only the poles
at $\sum_{i=1}^l m_i=-m, -m-1, -m-2, ...$, which correspond to massive
(oscillator) intermediate states in the channel $(T_{k_1}^{(+)}...
T_{k_l}^{(+)}T_{p_1}^{(-)}...T_{p_m}^{(-)})$, i.e. when a subset
of the right moving tachyons collide with {\it all} the left moving
ones. The fact that oscillator states can occur in this channel
is due to the fact that $\sum_{i=1}^m m(p_i)=2-n$ is fixed by kinematics.
Thus we see that the origin of the nice expression \nm\ is in an intricate
structure of poles. It is also easy to show that all other
poles that seem naively to appear in \tachcor\ have in fact vanishing residues.
For example, it is clear that the residues of the poles in
$\sum_{i=a}^b m(p_i)$
($a \neq 1$ and $b \neq m$) are given by correlation function
\chir\ with $r > 1$ and therefore vanish.
It is amusing to check explicitly that the residues
of the various poles are consistent with \nm\ (this is particularly
easy for the tachyon poles). We will not do that here.

\noindent{}3) We mentioned above that the
simplicity of the correlation functions in closed 2d string theory
is due from the
continuum  point of view to decoupling of certain discrete
states with negative energy (see \DIIFK). A subset of those are the
tachyons \vertex\ with $m(k)=1,2,3,\cdots$. In \nm\ it seems that such tachyons
``almost'' decouple, due to the factor of $\prod_i{1\over\Gamma(1-m_i)}$.
This breaks down for the particles on the boundary
between the right and left moving tachyons in \tachcor, since
$F_n$ develops a pole then. For the other $m_i$
we still find that the negative energy discrete states decouple.
This is the origin of the relative simplicity of \nm.
The role of the above discrete states requires (both here and in the
closed case) a much better understanding.

\noindent{} 4) There is a well known relation
\ref\KLT{ H. Kawai, D. Lewellen and S. Tye, Nucl. Phys.
{\bf B269} (1986) 1.}\
between open and closed tree amplitudes in critical string theory.
The derivation still applies here, so we may borrow the results. Assuming
that given the closed amplitudes, the open ones are uniquely
determined (the converse is certainly true), one way to prove \nm\
would be to plug it as an ansatz into the relations of \KLT.
We have only done this for five point functions, for which the relation
of \KLT\ reads:
\eqn\kaw{\eqalign{
A_{\rm closed}(k_1,...,k_5)=&\ao(k_1,k_2,k_3,k_4,k_5)
\ao(k_2,k_1,k_4, k_3,k_5)s_{12}s_{34}\cr
+&\ao(k_1,k_3,k_2,k_4,k_5)\ao(k_3,k_1,
k_4,k_2,k_5)s_{13}s_{24}\cr}}
where $s_{ij}\equiv\sin\pi k_i\cdot k_j=-\sin\pi(m_i+m_j)$
if $i,j$ have the same chirality, and $s_{ij}=\sin\pi m_im_j$ if they
have opposite chirality. It is amusing to verify \kaw\ by plugging
in \nm, and the closed string results \DFK.

\noindent{}5) The structure we find is very different from the closed case,
however \nm\ is only part of the amplitude. Could it be that summing
over channels restores the simpler closed string structure? In general
the answer is no.
Due to the fact mentioned above, that the amplitudes \chir\ satisfy
$\ao(k_i)\propto\delta_{r,1}$, we have only to symmetrize \Anm\ in
$\{k_i\}$ and $\{p_i\}$ separately. The function $F_n$ \fun\ is therefore
replaced by:
\eqn\hn{H_n(k_1,k_2,...,k_n)=
\sum_\sigma\prod_{l=1}^{n-1}{1\over\sin\pi\sum_{i=1}^l
m_{\sigma(i)}}}
where the sum over $\sigma$ runs over permutations of $1,2,...,n$.
The total amplitude after summing over channels, has the form
\nm\ with $F_n\rightarrow H_n$.
We haven't been able to explicitly sum \hn\ in general.
For low point functions one does get relatively
simple formulae; e.g for $N=4,5$ we find (after summing over channels \hn):
\eqn\Nfour{\eqalign{
A_{+++-}(k_1,...k_4)=&\prod_{i=1}^3{\Gamma({m_i\over2})\over\Gamma({1-m_i
\over2})};\;\;\;\;\;
A_{++--}(k_1,...,k_4)=0\cr
A_{++++-}(k_1,...,k_5)=&\prod_{i=1}^4\Gamma(m_i)
\sum_{1=i<j}^4s_{ij}\cr}}
Note that
tachyons with $m_i=1,2,3,\cdots$ do not decouple after
summing all diagrams. The poles still occur as a function of $m_i$. This
{\it does not} generalize to higher point functions. In general, the expression
\hn\ contains many poles in $\sum_i m_i$.

An interesting property of the amplitudes \fun, \nm, \hn\ is that
they follow from a space-time action describing free fermions
(in 2d). Indeed, redefining: $T_k^{(\pm)}\rightarrow\Gamma(1-m_k)
T_k^{(\pm)}$, the amplitude \nm\ takes the form
$\ao^{(n,m)}(k_i,p_j)=F_n(k_i)F_m(p_j)$, with $F_n$ given by
\fun.
Consider now the generating functional:
\eqn\W{{\cal W}(T^{(\pm)})=\int D\psi D\psi^* D\bar\psi D\bar\psi^*
\exp\left[-\int d^2Xe^{\Phi\over2}{\cal L}(\psi, T)\right]}
with\foot{We are using a non standard mode expansion for $\psi$:
$\psi(X^0, X^1)=\int d^2k e^{{i\over2}k\cdot X}\psi_k$.}:
\eqn\Spsi{{\cal L}(\psi, T)=
\psi^*\sin({\pi\alpha_-}\partial^+)\psi+
\bar\psi^*\sin({\pi\alpha_+}\partial^-)\bar\psi
+T^{(+)}\bar\psi^*\bar\psi+T^{(-)}\psi^*\psi~,}
where $\partial^{\pm}=\partial_0 \pm i \partial_1$.
One can check that ${\cal W}$ satisfies:
\eqn\gener{{\delta^{n+m}({\cal W/Z})\over\delta T_{k_1}^{(+)}...
\delta T_{k_n}^{(+)}\delta T_{p_1}^{(-)}...\delta T_{p_m}^{(-)}}=
H_n(k_1,...,k_n)H_m(p_1,...,p_m)}
To make sense of the functional integral \W, we have to specify the
zero mode prescription, and the coupling between left and right moving
fermions.
The generating functional ${\cal W}$ is defined with all zero modes
soaked up except $\psi^*(\vec k=0)$, $\bar\psi^*(\vec k=0)$.
The functional ${\cal Z}(T^{(\pm)})$
in \gener\ is given by the same expression as \W\ with all the zero modes
soaked up. The division by ${\cal Z}$ in \gener\ removes disconnected
diagrams.
In order to obtain \gener\ one has also to introduce a contact term
coupling between left and right:
$\langle\bar\psi(\vec k)\psi(-\vec k-\vec Q)\rangle=\gamma$,
which in space-time gives: $\langle \bar\psi(X)\psi(Y)\rangle=
\gamma e^{{Q\over2}X^0}
\delta^2(X-Y)$. Such a contact term can be achieved for example by modifying
the Lagrangian \Spsi\ as follows:
${\cal L}\rightarrow {\cal L}+\gamma\bar\psi\sin({\pi\alpha_-}\partial^+)
\sin({\pi\alpha_+}\partial^-)\psi$. Contact terms are usually
adjustable in quantum field theory, but it is not
clear to us whether this is the case here as well.
In any case, it is amusing that the non-trivial
S -- matrix of the open string tachyons is due to the
choice of these contact terms; without the contact terms $(\gamma=0)$,
the S -- matrix would have been trivial.

In \DIIFK\ it was shown that the closed string amplitudes also
follow from a space-time action, however one which was significantly
different from  \W, \Spsi. The main qualitative difference
between the two is that here ``Liouville'' $(X^0)$ momentum
is (anomalously) conserved
in \W, while in the closed case the natural description
of the amplitudes involved a field theory in which one had to integrate
over intermediate Liouville momenta; as a result of that,
the propagator, instead of containing poles as here, actually
had cuts; the infinite number of irreducible vertices
of \DIIFK\ is presumably related to the same phenomenon.
We do not know what is the origin of this difference in the behavior
of the two sectors of the theory.

One application of the free fermion representation \W, \Spsi\
is the generalization of \nm\ to arbitrary (non bulk) amplitudes.
It is easy to introduce the boundary cosmological constant in \Spsi:
this corresponds to $T^{(+)}\rightarrow T^{(+)}+\rho e^{{\alpha_-
\over2}X^0+i\alpha_0 X^1}$:
\eqn\Lrho{{\cal L}_\rho(\psi, T)=
\psi^*\sin({\pi\alpha_-}\partial^+)\psi+
\bar\psi^*\sin({\pi\alpha_+}\partial^-)\bar\psi
+\rho e^{{\alpha_-\over2}X^0+i\alpha_0 X^1}\bar\psi^*\bar\psi
+T^{(+)}\bar\psi^*\bar\psi+T^{(-)}\psi^*\psi~,}
\W, \Lrho\ can be used to calculate general tachyon correlation functions
\gener\ -- the fermions are still free.
We will describe the
details of the free fermion calculations elsewhere.

The fermions in \W, \Spsi\ can be thought of as the quarks at the
ends of the open string. This suggests a simple generalization to the
case of $U(N)$ gauge symmetry; one introduces $4N$ fermions
$\psi_i, \psi_i^*, \bar\psi_i, \bar\psi_i^*$, $i=1,2,...,N$;
replacing the kinetic term in \Spsi\ by
$\psi^*_i\sin({\pi\alpha_-}\partial^+)\psi_i$, etc, we see that
the action has a $U(N)$ symmetry $\psi_i\rightarrow\lambda_{ij}\psi_j$;
$\psi_i^*\rightarrow\lambda^\dagger_{ki}\psi^*_k$, $\lambda^\dagger\lambda=1$.
The source term in \Spsi\ is replaced by:
$T_a^{(-)}\lambda^a_{ij}\psi_i^*\psi_j$, so that the amplitudes
\gener\ are multiplied by ${\rm tr}(\lambda^{a_1}\lambda^{a_2}\cdots
\lambda^{a_{n+m}})$, as appropriate for Chan Paton factors.

Quite unexpectedly from the point of view of \tachcor,
the propagators in \Spsi\ are periodic in momentum space; the $\bar\psi$
propagator $\sin({\pi\alpha_+}\partial^-)=
\sin{\pi\alpha_+\over2}(\beta+k)$
is invariant under $k_x\equiv\beta+k\rightarrow k_x+2\alpha_-$, and
similarly for the $\psi$ propagator,
$k_y\equiv\beta-k\rightarrow k_y+2\alpha_+$. It is natural to consider
the theory where we identify momenta which differ by a period. In this
case momentum space is a torus, while
the conjugate variables, $x={1\over2}(X^1-iX^0)$, $y=-{1\over2}(X^1
+iX^0)$ live on a lattice $L$, with lattice spacings
$\delta_x={\pi\alpha_+}$,
$\delta_y={\pi\alpha_-}$.
Replacing $\int d^2X\rightarrow\sum_{(x,y)\in L}$
in \W\ we see that the ``dilaton factor'' disappears:
$e^{{1\over2}QX^0-i\alpha_0X^1}=1$, and \W, \Spsi\ assume the form:
\eqn\newW{{\cal W}(T^{(\pm)})=\prod_{(x,y)\in L}
\int D\psi(x,y) ...D\bar\psi^*(x,y)
\exp\left[-\sum_{(x,y)\in L}{\cal L}(\psi, T)\right]}
\eqn\newL{\eqalign{
{\cal L}(\psi, T)=&
\psi^*(x,y)\left[\psi(x,y+\delta_y)-\psi(x,y-\delta_y)\right]+
\bar\psi^*(x,y)\left[\bar\psi(x+\delta_x,y)-\bar\psi(x-\delta_x,y)\right]\cr
+&T^{(+)}\bar\psi^*\bar\psi+T^{(-)}\psi^*\psi~,\cr}}

Note that the space-time lattice is obtained only after continuing
$X^0\rightarrow iX^0$. One can think of
$X^0$ as a Liouville field, in which case the space-time lattice arises
when Liouville is treated as a Feigin Fuchs field. Before the continuation
$X^0$ is a non-compact dimension, as in \Spsi.

It is not clear
what is the relation of our fermions to those of the closed string matrix
models \SURF\ -- \CONE, however the two should be closely related.
It is also not clear to us why and how the space-time
lattice $L$ arose here, or equivalently,
what is the origin of the periodicity of $F_n$ \fun\ in $k$ space.
The string theory under consideration seems to develop a minimal length.
This phenomenon is similar to the one discussed in
\ref\irk{I. Klebanov and L. Susskind, Nucl. Phys. {\bf B309} (1988) 175.}.
The propagator \Spsi\ is reminiscent of the one obtained by Friedan
\ref\DFR{D. Friedan, Nucl. Phys. {\bf B271} (1986) 540.} in critical
string field theory.
In both cases the zeroes of the propagator are related to degenerate
representations, but
the relation between the two should be
elucidated further. There are many other questions concerning \W\ -- \newL\
which we haven't discussed; in particular, the role of \Spsi\ for
loop amplitudes, the coupling to the closed sector, the role
of the massive discrete states, etc. These issues will be addressed
elsewhere. It would also be interesting to solve the matrix
models corresponding to these theories. This may shed additional light
on their structure.

Another interesting issue concerns non trivial solutions of the open
$+$ closed string equations of motion \action. In the closed
case, the $SL(2, R)/U(1)$ coset
\ref\WBH{E. Witten, Phys. Rev. {\bf D44} (1991) 314.}\
leads to propagation in a 2d black hole geometry.
We can imitate that procedure here and construct a solution of the
open string equations of motion, by putting the $SL(2,R)/U(1)$ CFT on the
boundary of the world sheet manifold $\partial\cal M$. Repeating the steps
of
\ref\BER{M. Bershadsky and D. Kutasov, Phys. Lett. {\bf 266B} (1991) 345.},
we find an exact solution of the equations of motion in Euclidean space-time
with action (here we restrict to the case $\alpha_0=0$, as in \BER):
\eqn\pert{\eqalign{
{\cal S}={{\cal S}_0}+
          & m\int_{\cal M}d^2 \xi
  (\partial X^0+i\gamma
\partial X^1) (\bar \partial X^0+i
\gamma\bar \partial X^1)
                        e^{-Q X^0} +\cr
                          & q\int_{\partial\cal M}d\xi(\partial X^0+i
\gamma \partial X^1)e^{-{Q\over2}X^0}\cr}}
where ${\cal S}_0$ is the action in the trivial vacuum and
$\gamma=\sqrt{Q^2+1}$.
Comparing \pert\ to \action\ we see that we have found a solution
with linear dilaton $\Phi=Q X^0$, black hole background metric and
electric field
$E\simeq qe^{-{Q\over2}X^0}$. When the mass of the black hole $m$
vanishes, we have a flat $1+1$ dimensional space with an $X^0$
dependent electric field. For non zero mass of the black hole
the theory depends on the
parameter $m/{q^2}$ (in addition to the expectation
value of the dilaton), and the solution \pert\ describes
a charged black hole.
This solution is compatible
with
the lowest order equations of motion \eqmot: indeed, in 2d we have $F_{\mu\nu}
=\epsilon_{\mu\nu}E$, and \eqmot\ is equivalent to (see also
\ref\NAPPI{M. McGuigan, C. Nappi and S. Yost, IAS preprint
IASSNS-HEP-91/57 (1991).}):
$$\epsilon_{\mu\nu}\partial_\mu\left[e^{\Phi\over2}{E\over\sqrt{1+E^2}}
\right]=0$$
The solution has the form (for linear dilaton):
$E^{-2}=e^{QX^0}-1$. Recalling that \eqmot\ is only valid when $\delta E/E
<<1$ (which means here $e^{QX^0}>>{1\over1-Q}$, for $Q<1$), we finally find
in the region of validity of \eqmot, $E\simeq e^{-{Q\over2}X^0}$. It
is more complicated to find the exact solution this way, since the exact
form of \eqmot\ is not known.
The charged black hole solution suggested
here differs from the one discussed recently in the literature \ref\hor{N.
Ishibashi, M. Li and A. Steif, Santa Barbara preprint UCSBTH-91-28, (1991);
J. Horne and G. Horowitz, Santa Barbara preprint UCSBTH-91-39, (1991).}.
The solution \pert\ may be useful to study
pair production, and large electric field
behavior; it is interesting that at least from the low energy
Lagrangian giving \eqmot\ (the Born-Infeld Lagrangian ${\cal L}
=e^{\Phi\over2}\sqrt{{\rm det}(G+F)}$) it seems that physics
changes drastically in the region where the electric field exceeds
some critical value. This is reminiscent of the formation of a horizon
in the closed version of the theory, and should be studied further.
The free fermions should be very useful to study these issues.
There is a very natural way to couple the ``quarks'' to the gauge
field $A^\mu$, by replacing $\partial^{\pm}\rightarrow D^\pm=
\partial^\pm-iq A^\pm$ in \Spsi.
The action \Spsi\ is then invariant under the standard gauge
transformation, $\psi\rightarrow e^{iq\epsilon(X)}\psi$,
$A^\pm\rightarrow A^\pm-\partial^\pm\epsilon$. Tachyon
scattering in the background electric field of
\pert\ should then be given
by evaluating the free fermion path integral \W\ with modified
propagator, $\sin ({\pi\alpha_\mp}(\partial^\pm-iqA^\pm)$,
with $A^\pm\propto e^{-{Q\over2}X^0}$. The generalization to the non
abelian case is straightforward.
The properties of the theory in background fields will be left
for future work.

\bigbreak\bigskip\bigskip\centerline{{\bf Acknowledgements}}\nobreak

We would like to thank M. Douglas, E. Martinec,
C. Nappi, N. Seiberg, A. Tseytlin and C. Vafa
for valuable discussions. This work was partially supported by
DOE grant DE-AC02-76ER-03072 and
by Packard Fellowship 89/1624.

\listrefs

\end